\documentclass[12pt]{article}
\pdfoutput=1
\usepackage{hyperref}
\usepackage{cite}
\usepackage{color}
\usepackage{graphicx}
\usepackage{amsmath}
\usepackage{amssymb}
\usepackage{xspace}

\makeatletter
\@addtoreset{equation}{section}

\makeatletter
\renewcommand\section{\@startsection {section}{1}{\z@}%
                                   {-3.5ex \@plus -1ex \@minus -.2ex}
                                   {2.3ex \@plus.2ex}%
                                   {\normalfont\large\bfseries}}
\renewcommand\subsection{\@startsection{subsection}{2}{\z@}%
                                     {-3.25ex\@plus -1ex \@minus -.2ex}%
                                     {1.5ex \@plus .2ex}%
                                     {\normalfont\bfseries}}

\def\baselinestretch{1.2}
\newcommand{\be}{\begin{equation}}
\newcommand{\ee}{\end{equation}}
\newcommand{\beq}{\begin{eqnarray}}
\newcommand{\eeq}{\end{eqnarray}}

\newcommand{\gone}[1]{{}}

\begin{document}
\begin{titlepage}

\rule{0ex}{0ex}

\vfil

\begin{center}

{\bf \Large
Stability and boundedness in AdS/CFT\\ with double trace deformations
}

\vfil

Steven Casper$^1$, William Cottrell$^{1,2}$,  Akikazu Hashimoto$^1$,\\
 Andrew Loveridge$^1$, and Duncan Pettengill$^1$

\vfil

{}$^1$ Department of Physics, University of Wisconsin, Madison, WI 53706, USA

{}$^2$ Institute for Theoretical Physics Amsterdam, University of Amsterdam\\ 1098 XH Amsterdam, The Netherlands

\vfil

\end{center}

\begin{abstract}
Scalar fields on the bulk side of AdS/CFT correspondence can be
assigned unconventional boundary conditions, related to the
conventional one by Legendre transform. One can further perform double
trace deformations which relate the two boundary conditions via
renormalization group flow. Thinking of these operators as $S$ and $T$
transformations, respectively, we explore the $SL(2,{\bf R})$ family
of models which naively emerges from repeatedly applying these
operations. Depending on the parameters, the effective masses vary and
can render the theory unstable. However, unlike in the $SL(2,{\bf Z})$
structure previously seen in the context of vector fields in $AdS_4$,
some of the features arising from this exercise, such as the vacuum
susceptibility, turns out to be scheme dependent. We explain how
scheme independent physical content can be extracted in spite of some
degree of scheme dependence in certain quantities.
\end{abstract}
\vspace{0.5in}

\end{titlepage}
\renewcommand{\baselinestretch}{1.05}  

\section{Introduction}

In AdS/CFT correspondence, there is a well known dichotomy of
alternate boundary conditions originally pointed out by Klebanov and
Witten in \cite{Klebanov:1999tb}. The main idea is that the formula
for the scaling dimension of scalar operators dual to a bulk scalar of
mass $m$ in AdS/CFT correspondence
\be \Delta_+ = {d + \sqrt{d^2 + 4 m^2} \over 2} \label{DelM} \ee
also makes sense when one assigns the other branch of the square root
\be \Delta_- = {d - \sqrt{d^2 + 4 m^2} \over 2} \ . \ee
On the CFT side, the different assignment of dimensions corresponds to
defining a different CFT with different operator content. On the bulk
side, this dichotomy was interpreted as choice of boundary
condition. In the convention where the AdS metric takes the form
\be ds^2 = {r^2 \over R^2} (dt^2 + d \vec x^2) + {R^2 \over r^2} dr^2 \ee
the scalar fields behave asymptotically as
\be \phi(r) = P_1(\vec x) r^{-\Delta_-} (1 + \ldots) - P_2(\vec x)( r^{-\Delta_+} + \ldots)  \ . \ee
The standard convention is to fix the coefficient of the dominant
term, $P_1(\vec x)$, at the boundary and to infer the expectation
value of the dual operator from the value of $P_2(\vec x)$. This is
the branch where the dimension of operator is $\Delta_+$, and is
commonly referred to as the Dirichlet boundary condition for the
scalars in anti de Sitter space. In this scheme, $P_1(\vec x)$
interpreted as introducing a source term
\be \exp \left[ \int d^d x \ P_1(\vec x) {\cal O}(\vec x) \right] \ee
to the CFT path integral. 

In the alternate scheme, one fixes the subdomiant coefficient
$P_2(\vec x)$ and read off the expectation value of the operator of
dimension $\Delta_-$ from $P_1(\vec x)$. The role of $P_1$ and $P_2$
is therefore reversed, and this scheme is referred to as the Neumann
scheme.

These two distinct theories were further shown to be related by
Legendre transform of the sources \cite{Klebanov:1999tb}. This can be
understood as arising naturally on the bulk side as Legendre transform
which interchanges the boundary conditions. Klebanov and Witten showed
that this leads to expected mapping of the dimensions of operators
\cite{Klebanov:1999tb}.

The two theories are also related by renormalization group flow
induced by double trace deformation
\cite{Klebanov:1999tb,Witten:2001ua}. This flow can be visualized as
having the Neumann theory as the ultraviolet fixed point, deformed by
a double trace operator
\be \int d^d x  \ \alpha {\cal O}(\vec x)^2 \ee
whose dimension is
\be [ {\cal O}(\vec x)^2 ] = 2 \Delta_- \ee
and so is relevant for
\be 2 \Delta_- < d \ee
which is equivalent to the Breitenlohner-Freedman bound, which,
combined with the the unitarity bound, leads to constraint
\be -{d^2 \over 4} < m^2 <  -{d^2 \over 4} +1 . \ee
In the infra-red, the flow approaches the Dirichlet fixed point. The
double trace deformation is treatable using Hubbard-Stratonovich
techniques. The central charges at both the ultraviolet and the
infrared fixed points can be computed and was found to decrease along
the flow, as expected, at order $N^0$
\cite{Gubser:2002vv,Hartman:2006dy}. One can also compute correlation
functions and infer the cross-over between $\Delta_-$ scaling in the
ultraviolet and the $\Delta_+$ scaling in the infrared
\cite{Gubser:2002vv,Hartman:2006dy}. The double trace deformation
therefore introduces a continuous parameter which interpolates between
the Dirichlet and the Neumann theories.

So we have introduced two operations, the Legendre transform and the
double trace deformation, which acts on the space of theories. This
naturally leads one to wonder how these operations act in
combination. If one acts with Legendre transform, followed by a double
trace deformation, and then again with Legendre transform, is the
resulting theory equivalent, or distinct, from purely performing a
double trace deformation? In other words, does the double trace
deformation and Legendre transform commute? Related to these issues
is the question of how one parameterizes the space of theories on
which these transformations act.

Closely related issue was considered in the context of bulk vector
fields which are dual to current like operator by Witten in
\cite{Witten:2003ya}. In the setup of Witten, the dimension of bulk
was four, and the double trace deformation was related to Chern-Simons
term and as a result had quantized coefficients. The act of increasing
the Chern-Simons level acted as a discrete $T$-transformation while
the Legendre transform acted as $S$-transform, giving rise to an
$SL(2,{\bf Z})$ group of transformations. The set of theories then
corresponded to the set of $SL(2,{\bf Z})$ elements itself, on which
the group of transformations acted by left multiplication.

In this article, we will consider the scalar version of the double
trace deformation where the analogue of $S$ and $T$ transformation
exists, but where the $T$ transformation is continuous leading to the
$SL(2,{\bf R})$ family of theories. On first pass, we will find that
the $S$ conjugate of $T$ transformation turns out to be a contact-term
which parameterizes scheme dependence. This raises some puzzle
regarding the physicality/scheme independence of correlation functions
and equations of state. We will explain that scalar correlation
functions are indeed scheme dependent, but in such a way that some
observables, such as the onset of phase transitions, latent heat, and
critical scaling dimensions near a second order phase transitions are
scheme independent.

\section{$SL(2,R)$ transform of boundary condition and correlation functions}

Let us go back to the discussion of scalars, and consider a setup where the bulk geometry is an Eucledian
$AdS_{d+1}$-Schwarzschild geometry
\be ds^2 = {r^2 \over R^2} \left( f(r) dt^2 + d \vec x^2 \right) + {R^2 \over f(r) r^2} dr^2 \ee
with
\be f = 1-{r_0^d \over r^d} \ . \ee
Regularity requires the $t$ coordinate to be periodic
\be t = t + {1 \over T}, \qquad T = { d  \over 4 \pi R^2} r_0 \ . \ee
We will take $\vec x$ to be the coordinates in $(d-1)$ dimensions
which we also take to be living on ${\bf T}^{d-1}$ of volume $V_{d-1}$
but whose radii is much larger than $T^{-1}$. 

The reason for introducing thermal periodic boundary condition is to
regulate some observables in the infrared. It also allows various
observables to be interpreted in the context of thermal field
theory. Of course, one can just as easily take the zero temperature
limit in the end and see how the observables behave if desired.

We consider minimally coupled scalar in this background whose action is
\be S =  \int d^d x \int  dr \, \sqrt{g} \left({1\over 2} g^{rr}  (\partial_r \Phi)^2+{1\over 2} g^{xx}  (\partial_x \Phi)^2 + {1 \over 2} {m^2 \over R^2} \Phi^2 \right)  \ee 
where
\be \det g = \left({r^2 \over R^2}\right)^{d-1} \ .  \ee
and $m$ is dimensionless to match the convention used in (\ref{DelM}). 

We will now concentrate, for time being, on the zero momentum
component of the scalar field in the $\vec x$ directions, although we
will bring the momentum dependence later. Zero momentum component is
sufficient for discussing observables such as vacuum
susceptibility. It is also convenient to scale out dimensionful
quantities by setting
\be r=r_0 u  \label{r0}\ee
and
\be \Phi(r) =  \left({r_0 \over R^2}\right)^{-(d-1)/2} \phi(u) \label{phiPhi} \ . \ee

Then, the action becomes
\be S = {\cal N}  \int du\ u^{d-1} \left( {1 \over 2} f  u^2 (\partial_u \phi(u))^2 + {1 \over 2} m^2 \phi(u)^2  \right) \label{action} \ee
with
\be f=1- u^{-d} \ee
and
\be {\cal N} = {r_0 V_{d-1} \over   R^{d+1} T} \  . \label{calN1} \ee

In terms of this parametrization, we formulate the generating function
$Z[J]$ for scalar operators sourced by $J$ as follows.
\beq 
Z[J] &=& \int D \phi \exp \left[ -{\cal N}   \int_1^{u_c} du\  u^{d-1} \left( {1 \over 2} f u^2(\partial_u \phi(u))^2 + {1 \over 2} m^2 \phi(u)^2  \right)\right.
  \cr && \qquad  - {{\cal N} \over 2}  \Delta_- u_c^d \phi(u)^2 \label{ZJ} \\
  &&  \qquad \left.\left. \rule{0ex}{3ex}
  + {\cal N} (\Delta_+ - \Delta_-) \left( \alpha u_c^{2 \Delta_-} \phi(u_c)^2 + u_c^{\Delta_-}\beta \phi(u_c) J  \right)\right|_{u_{c} \rightarrow \infty}  \right] \nonumber \ .  \eeq
The first line of (\ref{ZJ}) is the action integrated up to the
ultraviolet cut-off at $u = u_c$ where $u_c$ is eventually taken to
infinity. The second line is the standard holographic renormalization
counter-term \cite{Bianchi:2001kw}. The third term includes terms
proportional to dimensionless parameters $\alpha$ and $\beta$, as well
as some dimensionless factor ${\cal N}(\Delta_+ - \Delta_-)$ which is
included for convenience.  They can be absorbed into $\alpha$ and
$\beta$ but we find it convenient to separate these factors as we did
in (\ref{ZJ}) for notional purposes. The term proportional to $\alpha$ is
recognizable as the double trace term, e.g.\ the term proportional to
$\lambda$ in equation (3.1) of \cite{Andrade:2011dg}.

The third line is also a boundary
term specifying the boundary condition. In terms of asymptotic
expansion
\be \phi(u) = p_1 u^{-\Delta_-} \left(1 + {\cal O}(u^{-1})\right) - p_2 u^{-\Delta_+} \left( 1 + {\cal O}(u^{-1}) \right) \label{p1p2} \ee
the Euler-Lagrange variation at $u=u_c$ in the large $u_c$ limit imposes the condition
\be
J = -{ 2  \alpha  \over \beta} p_1 + {1 \over \beta} p_2 \ . 
\ee
So, if $\alpha=0$, the boundary condition is Neumann. Turning on $\alpha$
corresponds then to the double trace deformation as was the case in
the treatment of \cite{Gubser:2002vv,Hartman:2006dy}. The operator
sourced by $J$ can also be identified as
\be {\cal O} \equiv
{1  \over (\Delta_+ - \Delta_-) {\cal N}} {\delta \over \delta J}   =  \beta  p_1 \ . 
\ee
This is precisely the prescription for reading off the expectation
value in Neumann theories.  The natural value to assign to $\beta$ is
$\beta=1$. Changing $\beta$ merely affects various normalization
conventions.

The issue we wish to explore is how this generating function further
transforms under Legendre transform and double trace deformation. We
will approach this question by making an educated guess for the
answer, and then subjecting the guess to tests.

The ansatz we wish to offer is to write the generating function in the
following form
\beq 
Z[J] &=& \int D \phi \exp \left[ -{\cal N}   \int_1^{u_c} du\  u^{d-1} \left( {1 \over 2} f u^2(\partial_u \phi(u))^2 + {1 \over 2} m^2 \phi(u)^2  \right)\right.
  \cr && \qquad  - {{\cal N} \over 2}  \Delta_- u_c^d \phi(u)^2 \label{ZJ2} \\
  &&  \qquad \left.\left. \rule{0ex}{3ex}
  + {\cal N} (\Delta_+ - \Delta_-) \left( \alpha u_c^{2 \Delta_-} \phi(u_c)^2 + u_c^{\Delta_-}\beta \phi(u_c) J  + \gamma J^2\right)\right|_{u_{c} \rightarrow \infty}  \right] \nonumber \ . \eeq
The term proportional to $\gamma$ is the new term, and is introduced to close the $SL(2,{\bf R})$ operation. It is the unique term which is missing in the bi-linears of $J$ and $\phi(u_c)$.   One can  immediately infer with this ansatz that the expectation value and the boundary condition takes the form
\beq
{\cal O} \equiv {1  \over (\Delta_+ - \Delta_-) {\cal N}} {\delta \over \delta J}  & = & \left(\beta  
-{4 \alpha  \gamma \over \beta}\right) p_1 + {2\gamma \over \beta} p_2 \label{NdJ} \\
J &=& -{ 2  \alpha  \over \beta} p_1 + {1 \over \beta} p_2  \label{JJ} \ . 
\eeq
which leads one to naturally parameterize $(\alpha, \beta, \gamma)$ in
terms of $SL(2,{\bf R})$ matrix
\be\left(\begin{array}{cc} a & b \\ c & d\end{array}\right) =
\left(\begin{array}{cc}        \beta  
     -{ 4 \alpha  \gamma \over \beta}&  {2\gamma \over \beta} \\
 -{2 \alpha  \over \beta}  & {1 \over \beta}\end{array}\right) \label{sl2r} \ee
and in this parametrization, setting $\gamma=1$ and $\beta=1$ gives
rise naturally to
\be\left(\begin{array}{cc} a & b \\ c & d\end{array}\right) =
\left(\begin{array}{cc}        1
& 0\\
-2 \alpha  & 1\end{array}\right) \ee
which is clearly the $T$ element of $SL(2,{\bf R})$. One can also
verify that starting with some generic values of $(\alpha, \beta,
\gamma)$ and performing Legendre transform gives rise to new set of
$(\alpha, \beta, \gamma)$ which corresponds to acting on $SL(2,{\bf
  R})$ parametrization by left multiplication by the $S$ element. As
such, we can conclude that (\ref{ZJ2}) is the general expression on
which the Legendre transform and double trace deformation can act.

Although we presented the analysis focusing on the zero momentum modes
along the boundary, it is straight forward to generalize the analysis
to non-zero modes, by writing
\beq Z[J] &=& \int D \phi \exp \left[ -{\cal N}  \sum_{\vec k}  \int_1^{u_c} du\  u^{d-1} \left( {1 \over 2} f u^2 \partial_u \phi_{\vec k}(u)  \partial_u \phi_{-\vec k}(u) + {1 \over 2} m^2 \phi_{\vec k}(u)\phi_{-\vec k}(u)  \right)\right.
   \cr && \qquad  - {{\cal N} \over 2}  \Delta_- u_c^d \phi_{\vec k}(u)\phi_{-\vec k}(u) \label{ZJ3} \\
  &&  \qquad \left.\left. \rule{0ex}{3ex}
  + {\cal N} (\Delta_+ - \Delta_-) \left( \alpha u_c^{2 \Delta_-} \phi_{\vec k}(u_c) \phi_{-\vec k} (u_c)+ u_c^{\Delta_-}\beta \phi_{\vec k}(u_c) J_{-\vec k}   + \gamma J_{\vec k} J_{-\vec k} \right)\right|_{u_{c} \rightarrow \infty}  \right] \nonumber \eeq
where the $d$ dimensional momentum label $\vec k$ is discrete since we
have compactified all spatial dimensions.

In this form, it is straightforward to compute the normalized two point function
\be G(\vec k) \equiv {1 \over (\Delta_+ - \Delta_-) {\cal N}} {\delta \over \delta J_{\vec k}} {\delta \over \delta J_{-\vec k}}  Z[J] = {a G_N(\vec k) + b \over c G_N(\vec k)  + d} \label{Glam} \ee
where
\be G_N(\vec k) = {d p_1(\vec k) \over d p_2(\vec k)} \ee
is the two point function for the Neumann boundary condition. For the
case of $d=2$ in the large volume limit for $\vec k$ entirely in the
spatial direction, this can be computed in closed form and takes the
form
\be G_N(k) = \frac{\Gamma (1+\nu) \Gamma \left(\frac{1}{2}
  (1-\nu -i k )\right) \Gamma \left(\frac{1}{2} (1-\nu+i k )\right)}{\Gamma (1-\nu ) \Gamma \left(\frac{1}{2} (1+\nu-i k)\right) \Gamma
   \left(\frac{1}{2} (1+\nu+i k)\right)}, \qquad \nu = \sqrt{1+m^2} \ . \label{chiNBTZ}
\ee
The values of $(a,b,c,d) $ are determined in terms of $(\alpha, \beta,
\gamma)$ by (\ref{sl2r}). The expression (\ref{Glam}) shows that the
two point function for generic $(\alpha, \beta, \gamma)$ is
parameterized naturally in terms of $SL(2,{\bf R})$. We have also seen
how the boundary condition (\ref{JJ}) and the operator (\ref{NdJ}) is
parameterized in terms of the $SL(2,{\bf R})$ data. So in a certain
sense, it is natural to think of the space of theories as also being
parameterized in terms of $SL(2,{\bf R})$. This is the main result of
this article. The implication and the interpretation of this result is
discussed in the following section.

The group $SL(2,{\bf R})$ has three generators.  Aside from the double
trace and contact term deformations, there is a relatively trivial
deformation (when $\alpha = \gamma=0$)
\be \left(\begin{array}{cc} a & b \\ c & d \end{array}\right)=
\left(\begin{array}{cc}
\beta  & 0 \cr
0  &  \beta^{-1}
\end{array}\right) \ . \label{lefthyper} \ee
Left multiplication by this element essentially amounts to adjusting
the normalization of ${\cal O}$ and as such is physically trivial.
Although the group manifold of $SL(2,{\bf R})$ is $AdS_3$ and is three
dimensional, one might be tempted to collapse the parameter space from
three to two by quotienting the space of theories with respect to left
multiplication by this hyperbolic element of $SL(2,{\bf
  R})$. Unfortunately, this introduces a multi-component quotient
space exactly analogous to what one finds when constructing the BTZ
orbifold by modding out with a discrete choice of $\beta$'s
\cite{Hemming:2002kd}. We find it more convenient to include this
trivial scaling as part of the theory space whose geometry is
familiar.

\section{Discussions}

In the previous section, we presented a parametrization of generating
functional (\ref{ZJ2}) and (\ref{ZJ3}) with parameters
$(\alpha,\beta,\gamma)$ related to $SL(2,{\bf R})$ parameters by
(\ref{sl2r}) such that
\begin{itemize}
\item Setting the parameters
\be \left(\begin{array}{cc} a & b \\ c & d \end{array}\right)=
\left(\begin{array}{cc} 1 & 0 \\ 0 & 1 \end{array}\right)
\ee
corresponds to the Neumann theory.

\item Acting with $T$ deformation
\be \left(\begin{array}{cc} a & b \\ c & d \end{array}\right)=
\left(\begin{array}{cc} 1 & 0 \\ - 2 \alpha  & 1 \end{array}\right)
\ee
corresponds to turning on the double trace deformation

\item Acting with $S$ deformation.
\be \left(\begin{array}{cc} a & b \\ c & d \end{array}\right)=
\left(\begin{array}{cc} 0 & -1 \\ 1   & 0 \end{array}\right)
\ee
corresponds to Legendre transform. 
\end{itemize}

One can therefore interpret the generic model parameterized by
$SL(2,{\bf R})$ as arising from successive action of $T$ and $S$
transformations. These data manifest themselves in observables through
relations (\ref{NdJ}) and (\ref{JJ}). We will now comment on the
interpretation of these results.

First observation we wish to make is the fact that the dependence on
$\gamma$ in (\ref{ZJ2}) and (\ref{ZJ3}) is merely that of a contact
term. In other words, they do not contribute to the correlation
function in position space at finite separation. This can also be seen
by re-writing the two point function (\ref{Glam}) in the form
\beq G(\vec k) &=& {1 \over c(c G_N(k) + d)} + {a \over c} \qquad (c \ne 0 ) \\
G(\vec k) & = & {1 \over d^2} G_N(k) + {b \over d} \qquad \qquad  \  (c=0)\ .  \label{G2}
\eeq
From this expression and (\ref{sl2r}), we see that the pole structure
only depends on $\alpha$.  The dependence on $\gamma$ is only in the
constant additive term $a/c$.  This is also reflected in the fact that
the boundary condition (\ref{JJ}) does not depend $\gamma$ and as a
result, the spectrum of small fluctuations are insensitive to this
$\gamma$. So, although we have identified an $SL(2,{\bf R})$ family of
partition function (\ref{ZJ2}) and (\ref{ZJ3}), we conclude that its physical
manifestation is mostly trivial.

There is however one subtlety in the interpretation of the dependence
on $\gamma$. The vacuum susceptibility
\be \chi \equiv G(k=0) = {\cal N} (\Delta_+ - \Delta_-) \langle {\cal O}_{\vec k=0} {\cal O}_{\vec k=0} \rangle \label{vsus} \ee
which parameterizes the stability of the vacuum is expected positive
definite by fluctuation dissipation theorem and is manifest in the
form of the right hand side of (\ref{vsus}). Indeed, for the Neumann
theory, one finds $\chi$ to be a positive number of order
one.\footnote{For $d=1$, this can be inferred from (\ref{chiNBTZ}).}
However, for general $(\alpha, \beta,\gamma)$, $\chi$ will inherit the
non-trivial dependence on $\gamma$ through (\ref{Glam}).  For the
Dirichlet model one obtains by taking the Legendre transform of the
Neumann theory, this susceptibility will turn out to be negative
despite the fact that the theory is expected to be sensible. By
contrast, small deformation of Neumann theory by double trace
deformation (\ref{ZJ}), which is expected to flow to the Dirichlet
theory in the infrared, has a positive susceptibility. How does one
make sense of all of these facts?

The answer is simply that the vacuum susceptibility is dependent on
contact terms because contact terms contribute at all momenta
including the zero mode.  The contact term itself should be
considered as the artifact of renormalization scheme dependence. The
situation here is different from that which was considered in
\cite{Witten:2003ya} where the large gauge transformations constrained
the analogue of our counter-terms to take on discrete scheme
independent values.

It seems that for the scalars, vacuum susceptibility is a scheme
dependent observable and must be calibrated by some renormalization
condition. This then suggests that susceptibility can not be treated
as an unambiguous (scheme independent) physical feature. If one
accepts this, however, the meaning of holographic second order phase
transitions such as
\cite{Gubser:2005ih,Hartnoll:2008vx,Hartnoll:2008kx}, whose salient
feature is the vanishing of vacuum susceptibility, all of a sudden
becomes mysterious.  How can one probe the behavior of vacuum
susceptibility switching signs when vacuum susceptibility itself is
scheme dependent?

To answer this question, we need to recall the fact that the momentum
expansion of the effective action for the order parameter \cite{Barbon:2002xk}
\be v_{\vec k} = \langle{\cal O}_{\vec k}\rangle \ee
is given by the inverse of the two point function\footnote{Here, we have set the momentum along the thermal circle to zero to keep the expression simple.} 
\be G(k)^{-1} = c_0 + c_2 k^2 + c_4 k^4 +  \ldots \label{sus}  \ . \ee
For the setup based on thermal AdS/CFT under consideration, one
expects all of these coefficients to generically be of order one,
which in our parametrization scheme (\ref{r0}) corresponds to the
scale being set by the temperature. (We are assuming that the radius
of the spatial coordinates is much larger than the radius of the
thermal circle.)  However, if for some special choice of parameters
$(\alpha, \beta, \gamma)$ one finds
\be c_0 \ll c_2 \sim c_4 \sim  \ldots \ee
then one finds a light effective degree of freedom whose mass is of order
\be m^2 = {c_0 \over c_2} \ee
in units set by the temperature. So only when $c_0$ happens to be
small, one expects an effective field theory description to be useful,
but this is precisely the regime near the second order phase
transition.

It is interesting to examine the behavior of the susceptibility and
the effective mass of our simple system (\ref{ZJ2}) starting with the
Neumann system with $\beta=1$, and gradually increasing $\alpha$,
keeping $\gamma$ at fixed zero. For the Neumann theory, one has
positive susceptibility and a positive effective mass squared of order
one in thermal units. As $\alpha$ is increased, however, the effective
mass decreases, until one reaches the point where
\be  \alpha = {1 \over 2 G_N(k=0)} \ . \ee
At this point, the system develops a tachyon and the susceptibility
flips sign. This is the tachyonic behavior previously discussed by
Troost in \cite{Troost:2003ig}. If this system is stabilized by
coupling to some other non-linear system, one expects to find a second
order phase transition around this point.

Furthermore, one sees in (\ref{sus}) that this is precisely when
susceptibility is going to infinity. But then since the susceptibility
is infinite, additive contribution from $\gamma$ is essentially
irrelevant. In other words, even though the specific values of the
susceptibility at generic point in the $SL(2,{\bf R})$ parameter space
is scheme dependent, the locus, and the behavior near, the critical
point is scheme independent.

To the extent that the contact term is affecting the operator
(\ref{NdJ}), one can think of the scheme dependence though contact
terms as field reparameterization ambiguity of the order parameter.
The reason that the effective field theory treatment works well
despite such ambiguity is the fact that the behavior near the critical
point is insensitive to such reparameterization ambiguities, and the
same principle is at work in the holographic setup. 

In essence, we are merely making a simple observation that effective
field theory analysis is reliable specifically when describing the
dynamics of some degree of freedom whose effective mass is much
lighter than the characteristic mass scale of the rest of the system. In
the context of $SL(2,{\bf R})$ family, we can reliably assess when the
condition for when such an effective description is valid, and infer
scheme independent universal features.

\begin{figure}
\centerline{\includegraphics[width=3in]{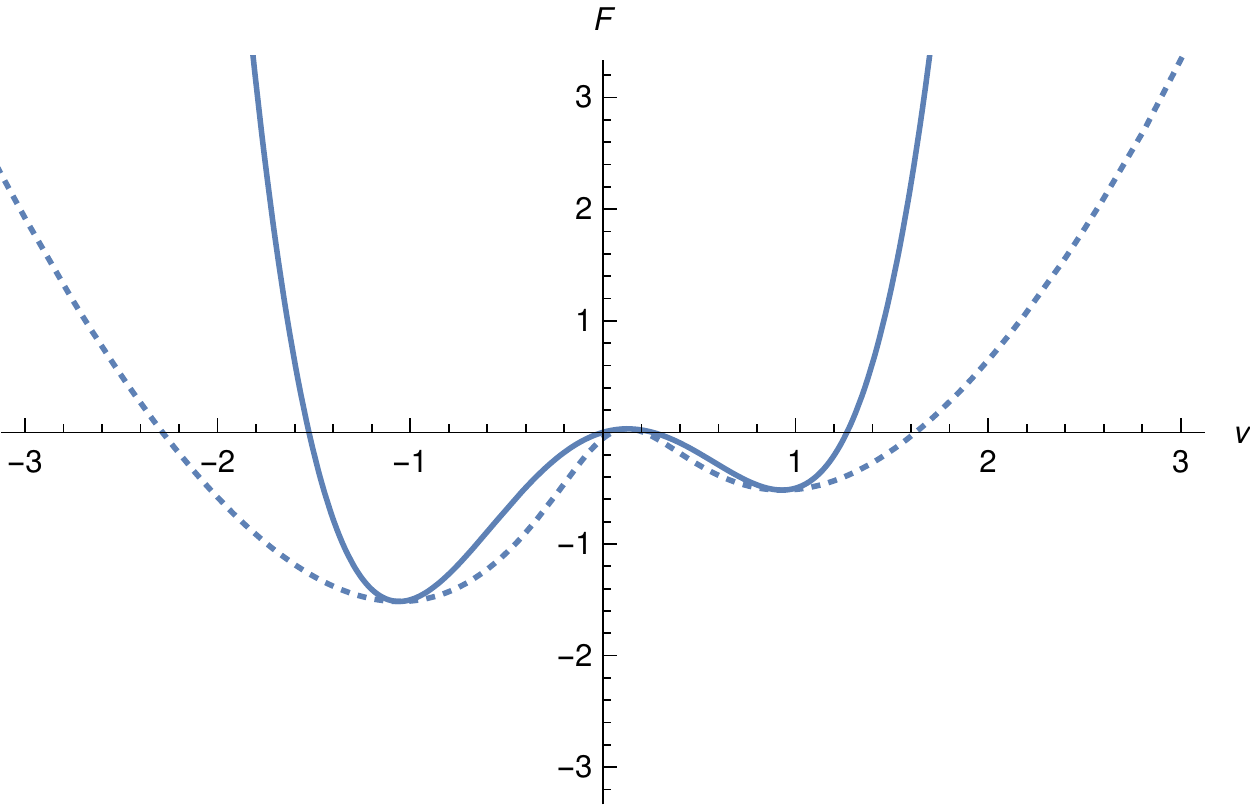}}
\caption{In this figure, we illustrate how a hypothetical free energy $F(v)$ (undotted) is transformed under field redefinition (\ref{NdJ}) of the order parameter $v' = v + \gamma {d F \over d v}$ (dotted) for some parameter $\gamma$. The form of the free energy changes, but the energies of the stationary point remains unchanged. Here, $p_1=v$ and $p_2= {dF \over dv}$. \label{figa}}
\end{figure}

An interesting issue to further explore is the extent to which global
stability issues such as Gibbs ruling of convex free energies (see
e.g. \cite{fisher}) is realized in light of some of these scheme
dependencies. Naively, there appears to be some tension between scheme
independence only for the near critical behavior of these systems,
which is local, in contrast to the constraints based on global
stability issues. Understanding this issue in a certain intersecting
brane system \cite{Cottrell:2015kra} was our original motivation to
explore these issues.  One can in fact see that coexistence of states
and the condition for first order phase transition to take place is
unaffected by the contact term deformations, as one would expect on
physical grounds, by the following argument.  Thinking of $p_2(p_1)$
as the equation of state with $p_1$ playing the role of order
parameter, the condition for two states to coexist at some fixed
external $\bar p_2$ is for there to be multiple $p_1$ satisfying the
condition
\be p_2(p_1^a)= p_2(p_1^b) = \bar p_2 \ . \ee
The difference in free energy for the stationary states at  $p_1=p_1^a$ and $p_1=p_1^b$ is then
\be \Delta F = \int_{p_1^a}^{p_1^b} d p_1 \, (p_2 - \bar p_2) \ . \ee
Now, the reparameterization of order parameter (\ref{NdJ}) transforms $p_1$ into $p_1 + \gamma p_2$. But then 
\be \Delta F' = \int_{p_1^a}^{p_1^b} (d p_1 + \gamma d p_2) (p_2 - \bar p_2) = \Delta F\ .\ee
The term proportional to $\gamma$ then drops out because the integral
starts and ends at the same value of $p_2$. So the $\Delta F$ after
the contact term deformation is unchanged. The equilibrium
configuration is the one with the lowest free energy. If there are
multiple degenerate global minimum, then the system can exist in
coexistence giving rise to a Gibbsian ruling.  So the precise form of
the free energy as a function of the order parameter remains scheme
dependent, but the possible existence of coexistence states and the
subsequent Gibbsian ruling is unaffected.  This point can be
illustrated, as we have done in figure \ref{figa}, by drawing how the
free energy transforms under field redefinition of the order
parameter. It would be interesting to further explore the consequences
of this observation in varoius holographic thermodynamic systems including \cite{Cottrell:2015kra}.

Before closing, let us further comment on the fact that the source
term $J$ appear non-linearly might seem unusual. This can be addressed
by introducing an auxiliary scalar field $\rho$ and writing
\beq 
Z[J] &=& \int D \phi \exp \left[ -{\cal N}   \int_1^{u_c} du\  u^{d-1} \left( {1 \over 2} f u^2(\partial_u \phi(u))^2 + {1 \over 2} m^2 \phi(u)^2  \right)\right.
  \cr && \qquad  - {{\cal N} \over 2}  \Delta_- u_c^d \phi(u)^2 \label{genZ4} \\
  &&  \qquad \left.\rule{0ex}{3ex}\left. 
  + {\cal N} (\Delta_+ - \Delta_-) \left( \alpha u_c^{2 \Delta_-} \phi(u_c)^2 + u_c^{\Delta_-}\beta (\phi(u_c)+\rho) J - {\beta^2 u_c^{2 \Delta_-} \over 4 \gamma} \rho^2 \right)\right|_{u_{c} \rightarrow \infty}  \right] \nonumber \ . \eeq
At this stage, we are not introducing kinetic term for the scalar
field $\rho$. Upon integrating out $\rho$ one immediately recovers
(\ref{ZJ2}).  One can however, prescribe a kinetic term and promote
$\rho$ into being a dynamical field. This is a potentially novel form
of boundary dynamics which may turn out to be interesting to further
explore.

\section*{Acknowledgements}

We would like to thank
A.~Buchel, D.~Marolf, and 
G.~Horowitz
for comments and discussions. This work is supported in part by the DOE grant DE-FG02-95ER40896.

\providecommand{\href}[2]{#2}\begingroup\raggedright\endgroup

\end{document}